# Computational Search for Novel Hard Chromium-Based Materials

*Alexander G. Kvashnin,[1,2,\*] Artem R. Oganov,[1,2,3,4] Artem I. Samtsevich,[1] Zahed Allahyari[1,2]*

[1] Skolkovo Institute of Science and Technology, Skolkovo Innovation Center 143026, 3 Nobel Street, Moscow, Russian Federation

[2] Moscow Institute of Physics and Technology, 141700, 9 Institutsky lane, Dolgoprudny, Russian Federation

[3] Department of Geosciences and Center for Materials by Design, Institute for Advanced Computational Science, State University of New York, Stony Brook, NY 11794-2100

[4] International Center for Materials Design, Northwestern Polytechnical University, Xi'an, 710072, China

**Abstract.**
Nitrides, carbides and borides of transition metals are an attractive class of hard materials. Our recent preliminary explorations of the binary chemical compounds indicated that chromium-based materials are among the hardest transition metal compounds. Motivated by this, here we explore in detail the binary Cr-B, Cr-C and Cr–N systems using global optimization techniques. Calculated enthalpy of formation and hardness of predicted materials were used for Pareto optimization to define the hardest materials with lowest energy. Our calculations recover all numerous known stable compounds (except $Cr_{23}C_6$ with its large unit cell) and discover a novel stable phase $Pmn2_1$-$Cr_2C$. We resolve the structure of $Cr_2N$ and find it to be of anti-$CaCl_2$ type (space group *Pnnm*). Many of these phases possess remarkable hardness, but only $CrB_4$ is superhard (Vickers hardness 48 GPa). Among chromium compounds, borides generally possess highest hardnesses and greatest stability. Under pressure, we predict stabilization of a TMDC-like phase of $Cr_2N$, a WC-type phase of CrN, and a new compound $CrN_4$. Nitrogen-rich chromium nitride $CrN_4$ is a high energy-density material featuring polymeric nitrogen chains. In the presence of metal atoms (e.g. Cr) polymerization of nitrogen takes place at much lower pressures: $CrN_4$ becomes stable at ~15 GPa (cf. 110 GPa for synthesis of pure polymeric nitrogen).

## Introduction

Generally, the hardest and most popular superhard materials known to date belong to two groups – (1) some B-C-N compounds and their derivatives (e.g., Refs. [1,2]), and (2) nitrides, carbides and borides of some transition metals. Compounds of the first class are semiconducting and brittle and the best known superhard phases (i.e. with Vickers hardness >40 GPa) belong to it, whereas those of the second class are metallic and more ductile. These two classes of very hard materials were uncovered in our preliminary computational searches. We explore a number of combinations with these elements, searching for materials with the best property (e.g. highest hardness, computed using the Lyakhov-Oganov model [3]). We indeed found diamond to be the hardest possible single crystal material, B-C-N phases to have the highest hardnesses, and among non-B-C-N compounds the Cr-B, Cr-C and Cr-N systems were indicated among the most promising for the existence of new hard and superhard materials. Indeed, recent theoretical studies of chromium nitrides and borides reported that $CrB_4$ and hypothetical metastable $CrN_2$ and can have hardness of 47 GPa,[4,5] and 46 GPa,[6] respectively.

Usually, chromium metal and its compounds are used in a wide range of applications mainly related to wear-resistant coatings,[7–12] cutting tools[13,14] and metal forming and plastic moulding applications.[15] Chromium nitride, CrN, is often used on medical implants and tools as a coating material due to its good wear, oxidation and corrosion resistance.[9–11] CrN is also a valuable component in advanced multicomponent coating systems, such as CrAlN, for hard, wear-resistant applications on cutting tools.[16]

Experimentally, six different chromium borides ($Cr_2B$, $Cr_5B_3$, CrB, $Cr_3B_4$, $CrB_2$ and $CrB_4$) are known,[17–21,4] and recently their mechanical characteristics were examined theoretically.[4,5,22] The experimental Vickers hardness of most Cr-B phases ranges from 20.7 to 24 GPa,[23,24] while Vickers hardness of $CrB_4$ phase was reported to be in a range of 29-44 GPa.[24]

It is known from experiments, that there are three stable chromium carbides, $Cr_{23}C_6$, $Cr_3C_2$ and $Cr_7C_3$.[12–14,25,26] Powders of $Cr_3C_2$ were prepared by heat-treatment of metastable chromium oxides of controlled morphology in $H_2$-$CH_4$ atmosphere.[8] Other metastable chromium carbides such as CrC and $Cr_3C$ have also been synthesized.[27–30] Theoretically calculated values of Vickers hardness of chromium carbide phases by Šimůnek model[31] vary from 13 to 32 GPa,[32] which is in a good agreement with experiments.[33–35]

Chromium nitrides are less studied, with most experimental works devoted to CrN and reporting the existence of a cubic paramagnetic B1-phase (NaCl-type) with $Fm\overline{3}m$ space group.[36,37] However, at temperatures below the Néel temperature (200-287 K)[37–40] B1-CrN phase transforms to an orthorhombic antiferromagnetic phase with $Pnma$ space group[37,39] and this transition was studied theoretically.[41] Today, electronic and magnetic properties of chromium nitride at low temperatures are actively studied.[37,41–44]

In addition to CrN, there is another stable compound $Cr_2N$, which appears together with CrN during the fabrication of Cr-N films and displays comparable wear resistance, but worse oxidation resistance.[45–48] Coating of $Cr_2N$ can be synthesized by either solid-state metathesis reaction of $CrCl_3$ with $Li_3N$[49] or by controlling the N flux.[47,48,50–52] Theoretically predicted crystal structure of $Cr_2N$[53] was based on experimental data made by Eriksson,[50] which reported about hexagonal close-packed structure with $P\overline{3}1m$ space group with lattice parameters $a = 4.752$ Å, $c = 4.429$ Å. Recently, comprehensive first-principles calculations of atomic structure and physical properties of different $Cr_2N$ phases with only varying distribution of the N atoms.[54]

It is important that none of the above mentioned works attempted global optimization of Cr-B, Cr-C and Cr-N systems and considered only already known or hypothesized compounds.

In this paper we explore the Cr-B, Cr-C and Cr-N systems using evolutionary structure prediction algorithm USPEX and density functional theory. The structure, stability, elastic constants and hardness of all considered phases are studied in detail.

**Results and Discussions**

First, we searched for stable compounds in the Cr-B, Cr-C and Cr-N systems at zero pressure. Based on the calculated enthalpies of formation of predicted phases for different compositions, convex hull diagrams were constructed, shown in Fig. 2a-c. Red points in the convex hull diagrams correspond to thermodynamically stable phases (see Fig. 2a-c), green points are studied metastable phases (see Fig. 2c,f). One can note from Fig. 2a, that five chromium borides were found including $I4/m$-$Cr_2B$, $I4/mcm$-$Cr_5B_3$, $Cmcm$-$CrB$, $Immm$-$Cr_3B_4$ $P6/mmm$-$CrB_2$ and $Pnnm$-$CrB_4$. All these predicted phases were already known from previous experimental works,[18,19,24] and are successfully found here in an unbiased calculation.

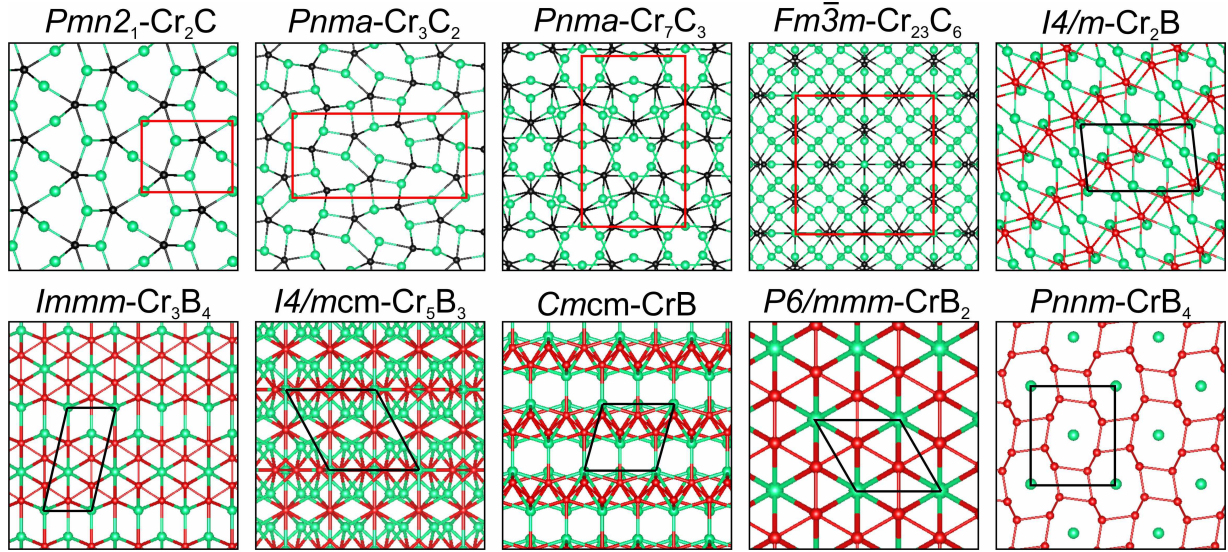

Fig. 1 Crystal structures of predicted Cr-C and Cr-B phases. Cr atoms are green, carbon is black, and boron is red.

During the evolutionary search of the Cr-C system, we found only three thermodynamically stable phases of chromium carbides shown in Fig. 2b by red points: $Pnma$-$Cr_7C_3$, $Pmn2_1$-$Cr_2C$, $Pnma$-$Cr_3C_2$. However, there is one stable phase $Cr_{23}C_6$ with $Fm\bar{3}m$ space group, which has not been found due to a large number of atoms (29) in the unit cell (blue point in Fig. 2b). The crystal structure of $Fm\bar{3}m$-$Cr_{23}C_6$ was taken from experiment [32] and the formation enthalpy was calculated to compare it with found structures. All found phases except $Pmn2$-$Cr_2C$ were synthesized experimentally.[8,25] In the Cr-N system, only two thermodynamically stable phases were found: $Pnma$-$CrN$ and $Pnnm$-$Cr_2N$. These phases were experimentally synthesized in a number of studies,[36,37,47–52] although the structure of $Cr_2N$ remained unknown. Other predicted phases, denoted by green points, are metastable (see Fig. 2 c). Structural parameters of all predicted phases are summarized in Table 1, and illustrated in Fig. 1.

Table 1. Details of atomic structure of predicted Cr-B and Cr-C phases.

| Comp. | Space group | Lattice parameters, Å | V, Å³/unit | ρ, g/cm³ |
|---|---|---|---|---|
| $Cr_2B$ | $I4/m$ | a = 4.21, b = 6.59, c = 4.04 | 27.95 | 6.82 |
| $Cr_5B_3$ | $I4/mcm$ | a = 5.43, b = 2.66, c = 4.56 | 73.15 | 6.64 |
| CrB | $Cmcm$ | a = 2.92, b = 7.84, c = 2.92 (theor: a = 2.93, b = 7.84, c = 2.92) [21] (exp: a = 2.959, b = 7.846, c = 2.919) [21] | 66.79 | 6.25 |
| $Cr_3B_4$ | $Immm$ | a = b = 2.92, c = 6.54 | 55.82 | 5.93 |
| $CrB_4$ | $Pnnm$ | a = 5.47, b = 2.85, c = 4.72 (exp: a = 5.48, b = 2.87, c = 4.74) [4,24] (exp: a = 5.48, b = 2.87, c = 4.75) [20] | 36.85 | 4.29 |
| CrB$_2$ | $P6/mmm$ | a = b = 2.98, c = 2.91<br>(theor: a = b = 2.97, c = 3.08) [55]<br>(exp: a = b = 2.97, c = 3.07) [17,24] | 22.46 | 5.44 |
| Cr$_7$C$_3$ | $Pnma$ | a = 4.48, b = 6.94, c = 12.01<br>(theor: a = 4.51, b = 6.91, c = 12.08) [32]<br>(exp: a = 4.53, b = 7.01, c = 12.14) [56] | 93.46 | 7.11 |
| Cr$_2$C | $Pmn2_1$ | a = 5.01, b = 2.82, c = 3.98 | 28.13 | 6.85 |
| Cr$_3$C$_2$ | $Pnma$ | a = 2.78, b = 5.47, c = 11.45<br>(theor: a = 2.79, b = 5.48, c = 11.47) [32]<br>(exp: a = 2.83, b = 5.55, c = 11.49) [57] | 43.72 | 6.84 |
| Cr$_{23}$C$_6$ | $Fm\bar{3}m$ | a = b = c = 10.82<br>(theor: a = b = c = 10.56) [32]<br>(exp: a = b = c = 10.66) [58] | 291.04 | 7.09 |

Let us now consider results of Pareto optimization shown in Fig. 2d-f. All points, which belong to a certain Pareto front, are connected by black line. The first Pareto front contains phases with simultaneously optimal high hardness (estimated using Lyakhov-Oganov model [3]) and maximum stability (measured as vertical distance from the convex hull). We consider the most promising phases, which are located mostly in the first five Pareto fronts, shown by red and open circles, which lie on the convex hull or close to it (see Fig. 2d-f). We note that the Lyakhov-Oganov model, convenient, numerically stable, and usually reliable, was used for Pareto-screening (and shown in Fig. 2) – however, it must be noted that Chen's model [59] is more accurate (these values are given in Table 2 and taken as final theoretical hardnesses in this work).

The most remarkable hardnesses, as well as largest negative enthalpies of formation, are seen in the Cr-B system. CrB$_4$ is predicted to be superhard ($H_v$ = 47.6 GPa), while all the other stable Cr-B phases display hardnesses below 35 GPa (see Table 2), which agrees well with reference experimental data. [23,24] Other phases with higher hardness have higher formation enthalpy and therefore are metastable or unstable at zero pressure. Most structures with hardness > 40 GPa are pure boron phases. Predicted stable Cr-C phases have Vickers hardness below 22 GPa (see Table 2), in agreement with experimental observations. [33–35] Phases with hardness about 70-80 GPa are hypothetical metastable carbon allotropes, and the hardest phase in the first Pareto front (Fig. 2e) with the hardness of 89 GPa is diamond with formation enthalpy of 0.028 eV/atom, which agrees well with reference data. [60,61]
Results of Pareto optimization of the Cr-N system show that thermodynamically stable CrN and Cr$_2$N phases display hardness up to 30 GPa. While metastable CrN$_2$ is predicted to be superhard using Gao's and Lyakhov-Oganov models, Chen's model gives a lower hardness (29.5 GPa). For metastable CrN$_4$ structures located in the first and second Pareto fronts (open circles in Fig. 2f) the predicted Lyakhov-Oganov hardness of ~ 60 GPa is a rare failing of this model: more accurate Chen's model predicts much lower values (see Table 2). For the other phases, agreement between different models of hardness is much better. We also calculated the ideal strength of $Pnma$-CrN, $P\bar{6}m2$-CrN, $Pnnm$-Cr$_2$N, $R3c$-CrN$_4$ and $Pnnm$-CrB$_4$ phases, to be equal to 38.2, 41.7, 37.3, 24.2 and 52.5 GPa, respectively. Ideal strength of $Pnnm$-CrB$_4$ was calculated before, [22] in close agreement with our result. Obtained values of ideal strength correspond well with data for Vickers hardness calculated by Chen's model.

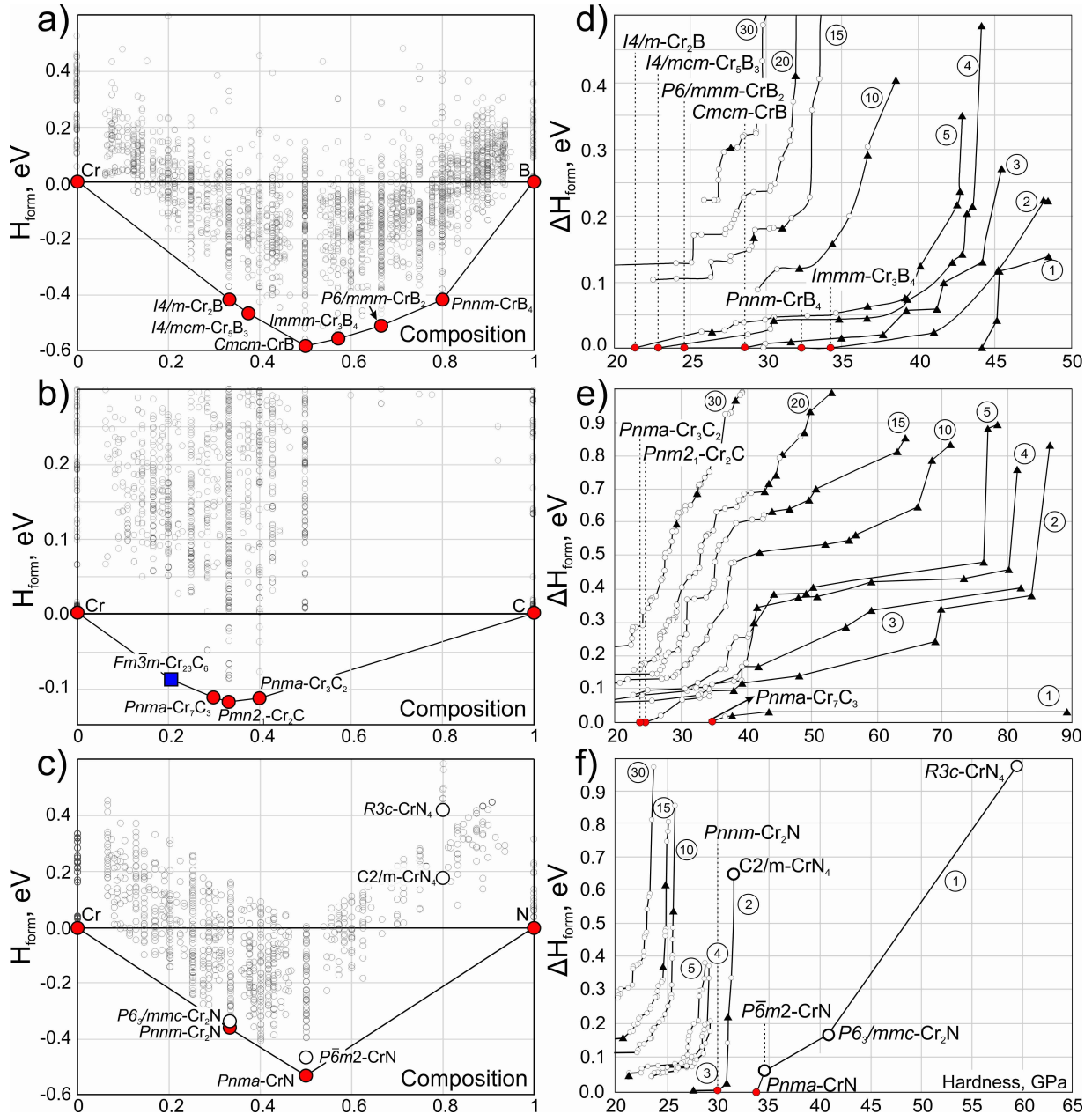

Fig. 2. Convex hull diagrams of a) Cr-B, b) Cr-C and c) Cr-N systems and results of Pareto optimization in terms of formation enthalpies and Vickers hardness, computed using the Lyakhov-Oganov model [3] for d) Cr-B, e) Cr-C and f) Cr-N systems. Numbers in circles denote the number of Pareto front. Full circles are stable, open circles - metastable binary phases, full triangles – one-component phases. Square is $Fm\bar{3}m-Cr_{23}C_6$ structure from Ref. [32]

We examined the mechanical properties of considered phases, summarized in Table 2. Considering the Cr-B system, the maximum value of bulk modulus was obtained for the $P6/mmm$-$CrB_2$ phase (278 GPa). The $Pnnm$-$CrB_4$ phase displays the largest value of shear modulus (252 GPa), which agrees extremely well with the theoretical and experimental values (267 [22] and 261 GPa, [4] respectively). Among chromium carbides, the highest bulk modulus is 296 GPa for $Pnma$-$Cr_3C_2$ and highest shear modulus is 292 GPa for $Pmn2_1$-$Cr_2C$ phase. The highest bulk modulus of Cr-N phases corresponds to $P\bar{6}m2$-CrN phase (312 GPa). It was expected that this WC-type phase would reveal exceptional mechanical properties (WC has bulk modulus of 439 GPa [62]). The bulk moduli of $Pnnm$ and $P6_3/mmc$ phases of $Cr_2N$ are 232 and 239 GPa, respectively. More detailed information on the elastic tensor of studied phases is summarized in Table S2 (Supporting Information).

Table 2. Mechanical properties of chromium-based materials. Bulk modulus (B), shear modulus (G), hardness calculated using Gao's model ($H_G$), Chen's model ($H_C$) and Lyakhov-Oganov model ($H_{LO}$), Pugh's modulus ratio ($k=G/B$) and thermal expansion for Cr-N phases at 300 K ($\alpha$).

| Comp. | Space group | B, GPa | G, GPa | $H_G$, GPa | $H_C$, GPa | $H_{LO}$, GPa | k | $\alpha$, $10^{-6}$ K$^{-1}$ |
|---|---|---|---|---|---|---|---|---|
| Cr$_2$B | $I4/m$ | 269.5 | 178.3 | 28.1 | 22.6 | 21.5 | 0.66 | – |
| Cr$_5$B$_3$ | $I4/mcm$ | 250.7 | 189.4 | 26.2 | 27.9 | 22.9 | 0.76 | – |
| CrB | $Cmcm$ | 255.3 (theor: 304.8) [21] (exp: 269) [21] | 209.5 (theor: 225.4) [21] | 32.6 | 33.2 | 28.6 | 0.82 | – |
| | | | | (exp: 19.2-23) [21,23] | | | | |
| Cr$_3$B$_4$ | $Immm$ | 276.6 | 202.8 | 32.9 | 28.1 | 34.1 | 0.73 | – |
| | | | | (exp: 20.9-23.0) [23] | | | | |
| CrB$_4$ | $Pnnm$ | 252.6 (theor: 265) [4] (exp: 232) [24] | 251.8 (theor: 267 [22]) (exp: 261 [4]) | 36.6 (theor: 46.8) [5] | 47.6 (theor: 48) [4] | 32.9 | 0.83 | – |
| | | | | (exp: 28.6-44) [24] | | | | |
| CrB$_2$ | $P6/mmm$ | 278.4 (theor: 298) [5] (exp: 228) [24] | 156.4 (theor: 172) [5] | 23.6 | 16.6 | 24.8 | 0.56 | – |
| | | | | (exp: 23.1-15.8) [24] | | | | |
| Cr$_7$C$_3$ | $Pnma$ | 264.6 (theor: 300.6) [32] | 104.4 (theor: 118) [32] | 25.1 (theor: 18.3 [32]) | 7.2 (exp: 16.9, [34] 17, [63] 16 [35]) | 33.1 | 0.44 | – |
| Cr$_2$C | $Pmn2_1$ | 292.8 | 184.5 | 27.3 | 21.6 | 24.5 | 0.63 | |
| Cr$_3$C$_2$ | $Pnma$ | 296.2 (theor: 312.9) [32] | 163.6 (theor: 162) [56] | 26.6 | 16.7 | 31.5 | 0.55 | – |
| | | | | (theor: 20.9 [32]) (exp: 18.9, [63] 18.3 [64]) | | | | |
| Cr$_{23}$C$_6$ Ref. [32] | $Fm\bar{3}m$ | 263.4 (theor: 282.3) [32] (exp: 300) [56] | 178.3 | 24.8 | 14.1 | 21.5 | 0.53 | – |
| | | | | (theor: 13.2, [32] exp: 15 [63]) | | | | |
| CrN (U-J=1 eV) | $Pnma$ | 221.4 (exp: 262) [39] | 152.1 | 35.8 | 21.4 | 34.8 | 0.72 | 2.01 |
| CrN (U-J=1 eV) | $P\bar{6}m2$ | 312.6 | 220.5 | 36.8 | 28.2 | 34.6 | 0.74 | 2.14 |
| Cr$_2$N | $Pnnm$ | 235.4 | 133.1 | 31.8 | 15.0 | 31.3 | 0.59 | 2.05 |
| Cr$_2$N | $P6_3/mmc$ | 239.8 | 116.1 | 37.9 | 11.0 | 41.0 | 0.47 | 2.55 |

| Comp. | Space group | $B$, GPa | $G$, GPa | $H_G$, GPa | $H_C$, GPa | $H_{LO}$, GPa | $k$ | $\alpha$, $10^{-6}$ K$^{-1}$ |
|---|---|---|---|---|---|---|---|---|
| CrN$_4$ ($U$-$J$=1 eV) | $C2/m$ | 26.7 | 21.8 | 46.8 | 2.2 | 31.6 | 0.82 | 6.52 |
| CrN$_4$ ($U$-$J$=1 eV) | $R3c$ | 176.6 | 101.1 | 57.2 | 12.5 | 59.5 | 0.59 | 1.39 |
| CrN$_2$ Ref. [6] | $P\bar{6}m2$ | 273.6 (theor: 366) [6] | 235.3 (theor: 256) [6] | 46.3 (theor: 45.9) [6] | 29.5 | 44.4 | 0.69 | – |

We paid more attention to the less studied Cr-N system and its stable and metastable phases. Part of the interest in new nitride phases comes from the possibility of reduction of the pressure of nitrogen polymerization for synthesis of high energy-density materials. It is necessary to compress pure nitrogen to >110 GPa [65] to obtain a polymeric phase, and such a high pressure precludes any practical applications. One of the possible ways to reduce the polymerization pressure is to combine nitrogen with metal ions (such as chromium, explored here). Indeed, it was found previously that presence of sodium reduces the polymerization pressure of nitrogen down to ~80 GPa in the compound NaN$_3$.[66]

In the convex hull diagram of the Cr-N system, three different composition of CrN$_x$ were found with x = 0.5, 1, 4. The X-ray diffraction (XRD) patterns are shown in Fig. 3 a. One can see good agreement between simulated and experimental [49] XRD patterns of *Pnma*-CrN, shown in the (i) panel of Fig. 3 a. We found that XRD pattern of the predicted *Pnnm*-Cr$_2$N agrees perfectly with experimental data from Ref. [49] (see Fig. 3 a, (ii) panel). This phase, observed in several experimental works,[47–52] remained structurally unresolved until now – but here we finally determine its crystal structure: *Pnnm* phase is isostructural to calcium chloride (CaCl$_2$) [67] and post-stishovite SiO$_2$ [68] (see Fig. 3 b).

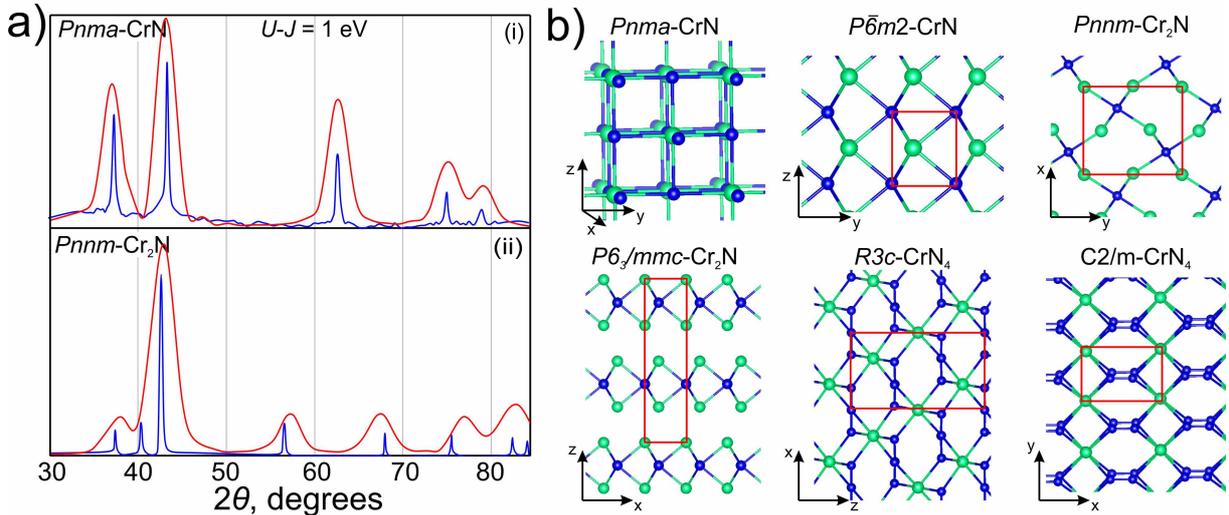

Fig. 3. a) Simulated X-ray diffraction pattern (XRD) with λ=1.54 Å. Blue lines are experimental XRD patterns from Ref. [49]; b) Crystal structures of CrN, Cr$_2$N and CrN$_4$ phases. Green spheres – Cr atoms, blue – N atoms.

*Pnma*-CrN phase has a NaCl-type structure with an orthorhombic distortion due to antiferromagnetic ordering, while predicted $\bar{6}m2$-CrN is isostructural to tungsten carbide (WC). Structural similarity suggests that $P\bar{6}m2$-CrN may have outstanding mechanical properties similar to those of WC. Another phase of Cr$_2$N with a space group $P6_3/mmc$ has layered structure and is isostructural to layered transition metal dichalcogenides (TMDCs), shown in Fig. 3b. This phase could be considered as a possible material for isolation of single layer of $Cr_2N$ using micromechanical cleavage. [69,70] Newly predicted $CrN_4$ is found in two forms, with space groups $R3c$ and $C2/m$; their structures are shown in Fig. 3b. Detailed structural parameters and energies above the convex hull (see Fig. 2c) of considered phases are summarized in Table 3.

Computed phonon densities of states for the $Pnma$ and $P\bar{6}m2$ CrN phases at zero pressure are shown in the (i) panel of Fig. 4 a and display the absence of imaginary phonon frequencies, which manifests about dynamical stability of both CrN phases. The phase transition pathway from $Pnma$ to $P\bar{6}m2$ CrN was modeled by the VCNEB method [71] and shown in Fig. S1 (see Supporting Information for details).

Table 3. Structural parameters of Cr-N phases.

| Comp. | Lattice parameters, Å | V, Å$^3$/unit | ρ, g/cm$^3$ | Positions | | | | $\Delta H_{form}$, eV |
|---|---|---|---|---|---|---|---|---|
| $Pnma$ CrN ($U$-$J$=1 eV) | a = c = 4.19, b = 4.17 (exp: a = 4.148 [37] a = 4.1513 [39]) | 18.33 | 5.97 | Cr N | 0.0 1/2 | 0.0 1/2 | 0.0 1/2 | 0.0 |
| $P\bar{6}m2$ CrN ($U$-$J$=1 eV) | a = b = 2.67, c = 2.59 | 16.05 | 6.82 | Cr N | 0.0 1/3 | 0.0 2/3 | 0.0 1/2 | 0.066 |
| $Pnnm$ $Cr_2N$ | a = 4.79, b = 4.33, c = 2.79 | 29.12 | 7.53 | Cr N | 0.164 0.0 | 0.242 1/2 | 0.0 1/2 | 0.0 |
| $P6_3/mmc$ $Cr_2N$ | a = b = 2.67, c = 9.19 | 28.29 | 6.93 | Cr N | 0.0 0.0 | -0.172 -0.162 | -0.112 1/4 | 0.005 |
| $C2/m$ $CrN_4$ ($U$-$J$=1 eV) | a = 7.64, b = 7.45, c = 3.91 | 44.41 | 3.32 | Cr Cr N N N | 0.0 0.0 1/2 0.285 -0.279 | 0.0 1/2 0.28 0.0 0.0 | 0.0 0.0 -0.349 0.371 0.328 | 0.364 |
| $R3c$ $CrN_4$ ($U$-$J$=1 eV) | a = b = 4.56, c = 13.81 | 49.74 | 4.18 | Cr N N | 0.0 0.339 0.0 | 0.0 0.376 0.0 | -0.157 1/4 -0.421 | 0.689 |
| $P\bar{6}m2$ $CrN_2$ Ref. [6] | a = b = 2.68, c = 3.67 (a = b = 2.72, c = 3.71) [6] | 22.76 (23.86) [6] | 4.24 | Cr N | 0.0 2/3 | 0.0 1/3 | 0.0 0.682 | 0.055 |

Both of the $Cr_2N$ phases (with space groups $Pnnm$ and $P6_3/mmc$) were found to be dynamically stable (see (ii) panel of Fig. 4a), the formed being energetically slightly more stable and matching perfectly the experimental XRD patterns (Fig. 3a). It is important to note that for metallic $Cr_2N$ phases we did not use the Hubbard $U$-term correction, in contrast to CrN and $CrN_4$ phases. Detailed information on the choice of $U$-$J$ parameter described in Supporting Information.

Two lowest-enthalpy CrN$_4$ phases that emerged from our evolutional searches are in fact high energy-density materials with polymeric nitrogen chains with 2 atoms ($C2/m$-CrN$_4$) and flat triangular NN$_3$-groups (similar to NO$_3$-groups, with oxygens replaced by nitrogens; $R3c$-CrN$_4$ is structurally similar to calcite CaCO$_3$ and NaNO$_3$) in the repeat unit. The effect of electron correlation is important in these phases: e.g., they are both dynamically unstable at $U$-$J$ = 0 eV (see Fig. S2), and dynamically stable with $U$-$J$ = 1 eV (see Fig. 4a, (iii) panel).

Both CrN$_4$ phases are metastable at zero pressure and even have positive enthalpies of formation (see Fig. 2c). However, at pressures above 5 GPa the formation enthalpy of $C2/m$ phase becomes negative (above 7.5 GPa for $R3c$-CrN$_4$), and at the pressure of 17 GPa the phase transition $C2/m \rightarrow R3c$ occurs. This means that $R3c$ phase of CrN$_4$ should be synthesizable under pressure more than 7 GPa. Calculations of phase transition pressure with $U$-$J$ from 0 to 5 eV gave the phase transition pressure in a region from 12 to 24 GPa at 0 K. At pressures above ~15 GPa CrN$_4$ becomes thermodynamically stable (see Fig. 5).

Containing polymeric nitrogen chains, at normal conditions CrN$_4$ can be a high energy-density material. We estimated the energy density of CrN$_4$ (equal to the enthalpy of reaction $CrN_4 \rightarrow CrN + 3/2\ N_2$) to be equal to 1.96 and 3.51 MJ/kg for $C2/m$ and $R3c$ phases, respectively. For comparison, the energy density of TNT (trinitrotoluene) is 4.6 MJ/kg,[72] for gunpowder 3 MJ/kg, for nitroglycerin 6.6 MJ/kg,[73] for lead azide 2.6 MJ/kg.[74] Our results show that the presence of metals (such as Cr) lowers the pressure of polymerization of nitrogen, even though with reduced (but still high) energy density.

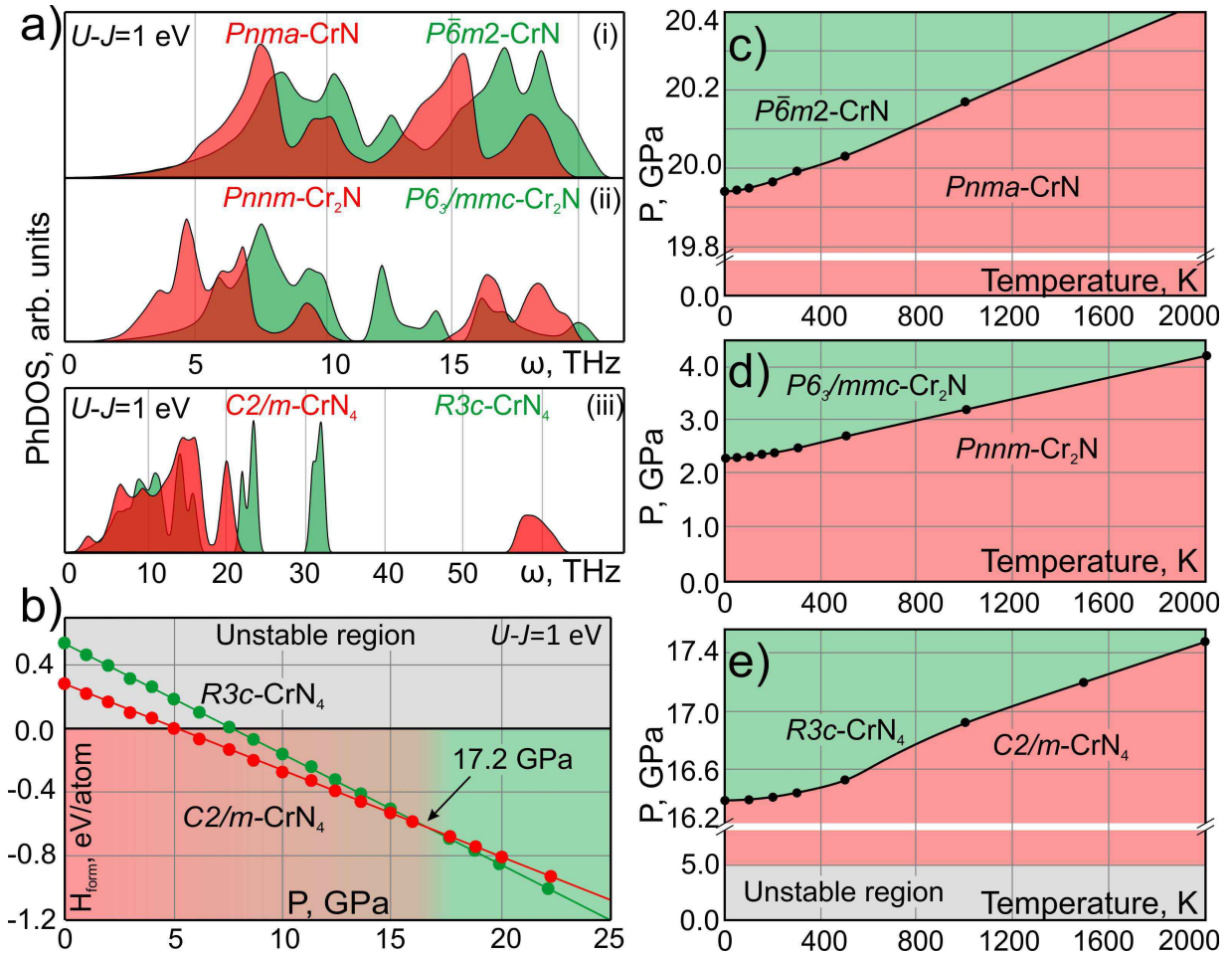

Fig. 4. a) Phonon densities of states of Cr-N phases; b) Dependence of the enthalpy of formation on the external pressure for CrN$_4$ phases. Phase diagrams of c) CrN, d) Cr$_2$N and e) CrN$_4$.

Conditions for experimental synthesis of CrN phases were estimated by computing phase diagrams, shown in Fig. 4c, where $Pnma \rightarrow P\bar{6}m2$ phase transition pressure at 0 K equals to 19.9 GPa, which is readily achievable in experiments. The phase boundary between $Pnnm$ and $P6_3/mmc$ phases of Cr$_2$N is shown in Fig. 4d, where $Pnnm$ phase undergoes phase transition to $P6_3/mmc$ under 2.2 GPa at 0 K. Thus, it should be possible to synthesize new Cr$_2$N phase with layered structure at very mild pressures, and this phase should remain dynamically stable upon decompression to ambient pressure. Computed phase diagram of the pressure-induced $C2/m \rightarrow R3c$ phase transition of CrN$_4$ is shown in Fig. 4 e, where the phase transition pressure equals to 16.4 GPa at 0 K.

The convex hull diagrams of Cr-N phases were calculated at the pressures of 10, 20 and 30 GPa as shown in Fig. 5. We see the same stable compositions as at zero pressure, and in addition CrN$_4$ becomes thermodynamically stable at pressures above ~15 GPa.

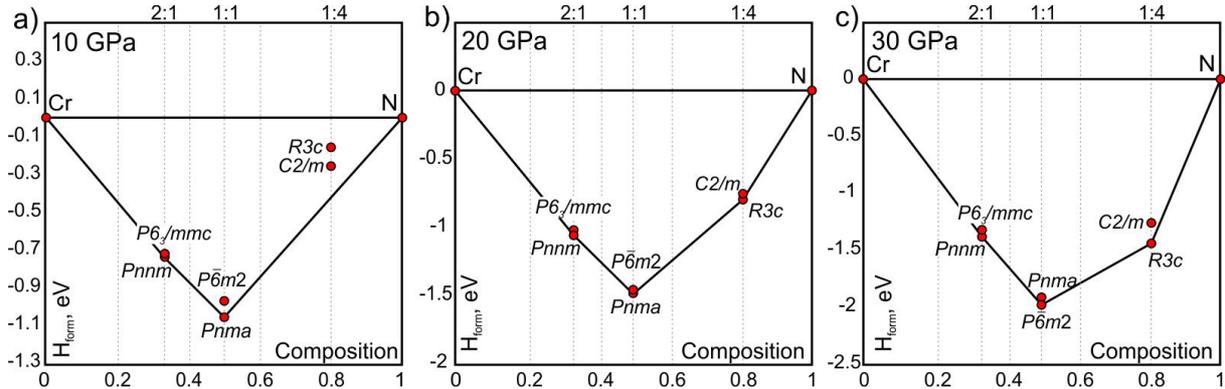

Fig. 5. Convex hull diagrams for Cr-N system at a) 10 GPa, b) 20 GPa, c) 30 GPa.

**Conclusions**

In this work, we studied new phases in the Cr-B, Cr-C and Cr-N systems using global optimization combined with Pareto optimization technique, which allows us to search for new stable materials with outstanding hardness. We found all experimentally known chromium borides, carbides and nitrides (except Cr$_{23}$C$_6$ with a relatively large unit cell) and predicted several new phases. Hardness of the predicted phases was calculated using different models and compared with available experimental and theoretical data. Overall, chromium borides are shown to possess highest hardnesses and largest negative enthalpies of formation, compared to carbides and nitrides. The only thermodynamically stable superhard compound here is CrB$_4$ with the predicted hardness of ~48 GPa, in excellent agreement with experiments.[24] Detailed investigation of the less studied Cr-N system was carried out. The previously unresolved crystal structure of Cr$_2$N was shown to be of anti-CaCl$_2$ type (space group $Pnnm$). We found that synthesis of CrN$_4$ phases with energy density up to 3 MJ/kg and featuring polymeric nitrogen chains can be realized by applying pressure above ~15 GPa, much lower than 110 GPa needed to synthesize pure polymeric nitrogen.

**Methods**

Stable phases in the Cr-B, Cr-C and Cr-N systems were predicted using first-principles variable-composition evolutionary algorithm (EA) in coupling with Pareto optimization technique as implemented in the USPEX code.[75–80] Here, evolutionary searches were combined with structure relaxations using density functional theory (DFT)[81,82] within the spin-polarized generalized gradient approximation (Perdew-Burke-Ernzerhof functional),[83] as implemented in the VASP [84–86] package. The plane–wave energy cutoff was set to 500 eV. For studying phase transition pathways of CrN phases, we used the variable-cell nudged elastic band method (VCNEB) [71] as implemented in the USPEX code. In order to take into account strong electron correlations between the localized 3$d$-electrons of Cr atoms, the GGA+$U$ approach within Dudarev's formulation [87,88] was applied in some cases (unless explicitly stated otherwise, $U-J = 0$ was used). For Brillouin zone sampling, Γ-centered $k$-meshes of 2π×0.05 Å$^{-1}$ resolution were used, ensuring the convergence of total energies to better than 10$^{-6}$ eV/atom. During structure searches, the first generation was produced randomly within 16 atoms in the unit cell, and succeeding generations were obtained by applying heredity (40%), softmutation (20%), transmutation (20%) operations, respectively and 20% of each generation was produced using random symmetry generator. Two types of variable-composition calculations were performed in each binary system (Cr-B, Cr-C, Cr-N): (1) optimizing stability and (2) jointly optimizing stability and hardness with Pareto ranking of all structures (in the latter case, the fitness of each structure was taken to be equal to the order of its Pareto front).

For the predicted crystal structures, we performed high-quality calculations of their physical properties. Crystal structures were relaxed until the maximum net force on atoms became less than 0.01 eV/Å. The Monkhorst–Pack scheme [89] was used to sample the Brillouin zone, using 12·12·12 (*Pnma*-CrN), 8·8·10 (*Pnnm*-Cr$_2$N), 12·12·8 (*C*2/*m*-CrN$_4$), 6·6·6 (*Fm$\bar{3}$m*-Cr$_{23}$C$_6$), 8·8·8 (*Pmn*2$_1$-Cr$_2$C), 8·6·4 (*Pnma*-Cr$_3$C$_2$), 8·6·4 (*Pnma*-Cr$_7$C$_3$), 8·8·8 (*I*4/*m*-Cr$_2$B), 8·8·6 (*Immm*-Cr$_3$B$_4$), 8·8·6 (*I*4/*mcm*-Cr$_5$B$_3$), 8·8·8 (*Cmcm*-CrB), 6·8·6 (*Pnnm*-CrB$_4$), while for hexagonal lattices the Γ-centered grid was used with $k$-points mesh of 12·12·12 (*P*6/*mmm*-CrB$_2$), 12·12·12 (*P$\bar{6}$m*2-CrN), 8·8·4 (*P*6$_3$/*mmc*-Cr$_2$N), 8·8·6 (*R*3*c*-CrN$_4$).

The hardness was estimated according to three models of hardness: Lyakhov-Oganov model [3] ($H_{LO}$), Gao's model [90] ($H_G$) and Chen's model [59] ($H_C$), in the latter hardness is calculated using the following relation:

$$H_C = 2 \cdot (k^2 \cdot G)^{0.585} - 3$$

where $k$ is the Pugh ratio ($k=G/B$), and $G$ is shear modulus and $B$ the bulk modulus. The bulk and shear moduli were calculated via Voigt-Reuss-Hill (VRH) averaging. [41]

The phase diagram was obtained using the computed Gibbs free energies $G$ of the relevant phases in the quasiharmonic approximation: [91]

$$G(P,T) = E_0(V) + F_{vib}(T,V) + P(T,V)V,$$

where $E_0$ is the total energy from the DFT calculations and $F_{vib}$ is vibrational Helmholtz free energy calculated from the following relation:

$$F_{vib}(T,V) = k_B T \int_\Omega g(\omega(V)) \ln\left[1 - \exp\left(-\frac{\hbar\omega(V)}{k_B T}\right)\right] d\omega + \frac{1}{2}\int g(\omega(V))\hbar\omega d\omega,$$

and pressure is

$$P(T,V) = -\frac{\partial(E_0(V) + F_{vib}(T,V))}{\partial V}.$$

Here $g(\omega(V))$ is the phonon density of states at the given pressure, calculated from forces on atoms with atomic finite displacements using density-functional perturbation theory (DFPT) implemented in the VASP package, [84–86] and the phonon frequencies are calculated from the force constants using the PHONOPY package. [92,93] Once Gibbs free energies are computed, phase equilibrium lines on the phase diagram are determined as loci of points where free energies of phases are equal. The chosen approach is validated by a number of reference papers [91,94–99] that calculated the phase diagram $P(T)$ of various materials. Crystal structures of predicted phases were generated using VESTA software. [100]

**Acknowledgements**

The work was supported by Russian Science Foundation (№ 16-13-10459). Calculations were performed on the Rurik supercomputer at MIPT. The authors thank Prof. Vladislav A. Blatov for help in the application of TOPOS package for design of the initial transition pathway in CrN (see Supporting Information).

**Supporting Information Available**: Detailed description of the mechanism of the phase transition of CrN from NaCl-type to WC-type structure. The details of calculations of Cr-N system with $DFT+U$ approach. Calculated elastic tensor of studied Cr-C, Cr-B and Cr-N systems compared with reference data. Electronic properties of Cr-N phases.